\documentclass{ws-procs961x669}            % default, citations in superscript
\usepackage{verbatim}
\usepackage{aas_macros}
\usepackage{hyperref}

\begin{document}
\title{Hydrodynamical transport of angular momentum in accretion disks in the presence of nonlinear perturbations due to 
noise}

\author{Subham Ghosh$^*$ and Banibrata Mukhopadhyay$^\dagger$}

\address{Department of Physics, Indian Institute of Science,\\
Bangalore, Karnataka, 560012, India\\
$^*$subham@iisc.ac.in\\
$^\dagger$bm@iisc.ac.in}

\begin{comment}
\author{A. N. Author}

\address{Group, Laboratory, Street,\\
City, State ZIP/Zone, Country\\
E-mail: an\_author@laboratory.com}
\end{comment}

\begin{abstract}
The origin of hydrodynamical instability and turbulence in the Keplerian accretion disk is a long-standing puzzle. The 
flow therein is linearly stable. Here we explore the evolution of perturbation in this flow in the presence of an 
additional force. Such a force, which is expected to be stochastic in nature hence behaving as noise, could result from 
thermal fluctuations (however small be), grain–fluid interactions, feedback from outflows in 
astrophysical disks, etc. We essentially establish the evolution of nonlinear perturbation in the presence of Coriolis and 
external forces, which is the modified Landau equation. We obtain that even in the linear regime, under suitable forcing 
and Reynolds number, the otherwise least stable perturbation evolves to a very large saturated amplitude, leading to 
nonlinearity and plausible turbulence. Hence, forcing essentially leads a linear stable mode to unstable. We further show 
that nonlinear perturbation diverges at a shorter time-scale in the presence of force, leading to a fast transition to 
turbulence. Interestingly, the emergence of nonlinearity depends only on the force but not on the initial amplitude of 
perturbation, unlike the original Landau equation-based solution.
\end{abstract}

\keywords{accretion, accretion disks -- hydrodynamics -- instabilities -- turbulence}

\bodymatter
\section{Introduction}
\label{sec:intro}
Origin of instability and plausible turbulence in Rayleigh stable flows, e.g., the Keplerian accretion flow, is a 
long-standing problem. While such flows are evident to be turbulent, they are linearly stable for any Reynolds number 
($Re$). Although people\cite{Shakura_1972, Lynden-Bell_1974} described the underlying flow in accretion disks to be 
turbulent with an effective turbulent viscosity, the origin of turbulence therein was not clear until Balbus and 
Hawley\cite{Balbus_1991} proposed the idea of magnetorotational instability (MRI) following Velikhov \cite{Velikhov_1959} 
and Chandrasekhar \cite{Chandrasekhar_1960}. MRI is a Magnetohydrodynamical (MHD) instability, and it operates due to the 
coupling between the weak magnetic field and the rotation of the fluid parcel. Although MRI is a popular instability 
mechanism, it poses several problems. There are astrophysical bodies where the ionization fraction is tiny as they 
are cold. In such systems, MRI gets suppressed \cite{Bai_2013_ApJ}. Apart from that, beyond a certain value of the 
toroidal component of the magnetic field for compressible plasma, MRI gets suppressed\cite{Pessah_2005, Das_2018}. There 
are several other cites too where MRI seizes to work or becomes suppressed\cite{Nath_2015, Bhatia_2016}. We, therefore, 
venture for a hydrodynamical origin of nonlinearity and hence plausible turbulence in the accretion disk. Our emphasis is 
the conventional linear instability when perturbation grows exponentially, unlike the case of transient growth. We 
particularly consider here an extra force \cite{Nath_2016}, to fulfill our purpose. The examples \cite{Ghosh_2020} of the 
origin of such force in astrophysical context, particularly in accretion disks, could be: the interaction between the dust 
grains and fluid parcel in protoplanetary disks, back reactions of outflow/jet to accretion disks.

\section{Formalism of the problem}
\label{sec:formalism_of_the_problem}

\begin{figure}
\begin{center}
\includegraphics[width=2.5in]{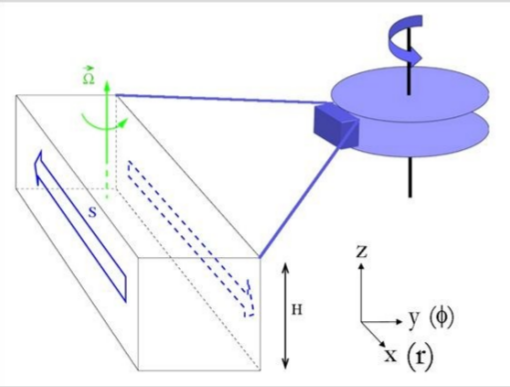}
\caption{Model picture of local cubical box in accretion disk where we perform the analysis 
(http://ipag.obs.ujf-grenoble.fr/~longarep/astrophysics.html). Within the box, the Cartesian coordinate $x$ is 
along the radial cylindrical coordinate $r$ (with respect to the center of the accretion disk), $y$ is 
along $\phi$, and $z$ is same in both the systems.}
\label{fig:rotating_couette_flow}
\end{center}
\end{figure}

To formulate the problem, we have considered a local box at a local patch in the accretion disk, as shown in the 
Fig.~\ref{fig:rotating_couette_flow}. Although the accretion disk possesses cylindrical geometry, In the local box, 
we 
study the motion of the local fluid element using the Cartesian coordinates. The one to one correspondence between the 
Cartesian ($x,y,z$) and cylindrical ($r,\phi, z$) coordinates are shown in the Fig.~\ref{fig:rotating_couette_flow}. The 
Cartesian coordinates $x,y,z$ are along the cylindrical coordinates $r,\phi, z$. The details of the local formulation are 
in Mukhopadhyay et al.\cite{man_2005, Mukho_2011}, and Ghosh and Mukhopadhyay\cite{Ghosh_2021}. As we 
confine ourselves in the local region of the accretion disks, the corresponding flow has been considered 
incompressible\cite{Yecko_2004, man_2005, amn_2005, Rincon_2007, Nath_2015}. The governing equations are the 
ensemble-averaged Orr-Sommerfeld and Squire equations in the presence of Coriolis force and an additional stochastic 
force\cite{Mukhopadhyay_2013, Ghosh_2020}, which is delta correlated  with nonzero mean. The Orr-Sommerfeld and Squire 
equations are obtained by eliminating the pressure term from the corresponding Navier-Stokes equations and utilizing the 
continuity equation for the incompressible flow. The governing equations are given by,
\begin{eqnarray}
 \left(\frac{\partial}{\partial t} + U\frac{\partial}{\partial y} \right)\nabla^2u - U''\frac{\partial u}{\partial y} 
+\frac{2}{q}\frac{\partial \zeta}{\partial z} - \frac{1}{Re}\nabla^4u + \Gamma_1 = NL^{u},
\label{eq:orr_sommerfeld_eq}
\end{eqnarray}

\begin{eqnarray}
\left(\frac{\partial}{\partial t} + U\frac{\partial}{\partial y} \right)\zeta - U'\frac{\partial u}{\partial z} 
-\frac{2}{q}\frac{\partial u}{\partial z} - \frac{1}{Re}\nabla^2\zeta + \Gamma_2  = NL^{\zeta},
\label{eq:squire_eq}
\end{eqnarray}
where $u$ and $\zeta$ are respectively the $x$-component of the velocity and vorticity perturbations after ensemble 
averaging, $U$ the $y$-component of background velocity which for the present purpose of plane shear in the 
dimensionless units is $-x$, $q$ the rotation parameter with $\Omega(r)\propto 1/r^q$, $\Omega(r)$ being the angular 
frequency of the fluid parcel at radius $r$, $\Gamma$-s are the corresponding constant means of stochastic forces (white 
noise with nonzero mean due to gravity making the system biased\cite{Nath_2016}) in the system described below, and 
$NL$-s are the non-linear terms due to the perturbation. The $x$-component of vorticity and the non-linear terms are 
given by
\begin{eqnarray}
 \zeta &=& \frac{\partial w}{\partial y} - \frac{\partial v}{\partial z},
 \label{eq:x_comp_vorti}
 \\
 NL^u &=& -\nabla^2\{(\textbf{u}'\cdot \nabla)u\}+\frac{\partial}{\partial x} \nabla \cdot \{(\textbf{u}'\cdot \nabla) 
\textbf{u}'\},
 \label{eq:non_orr_sommerfeld_eq}
 \\
 NL^{\zeta} &=& -\frac{\partial}{\partial y}\{(\textbf{u}'\cdot \nabla)w\} + \frac{\partial}{\partial 
z}\{(\textbf{u}'\cdot \nabla) v\},
 \label{eq:non_squire_eq}
\end{eqnarray}
where $\textbf{u}'=(u,v,w)$, which is the perturbed velocity vector. For the complete description of the flow, equations 
\ref{eq:orr_sommerfeld_eq} and \ref{eq:squire_eq} are supplemented with the equation of continuity, which is given by 
\begin{equation}
 \nabla \cdot \textbf{u}' = 0,
 \label{eq:continuity_eq}
\end{equation}
To solve the governing equations, we have no-slip boundary conditions along $x$- direction \cite{Ellingsen_1970, 
man_2005, Ghosh_2020}, i.e. $u = 
v = w = 0$ at $x=\pm1$ or equivalently
\begin{equation}
 u = \frac{\partial u}{\partial x} = \zeta = 0,\,\ {\rm at}\,\, x = \pm1.
 \label{eq:no-slip_bc}
\end{equation}

\subsection{Linear Theory}
\begin{figure}
\begin{center}
\includegraphics[width=3in]{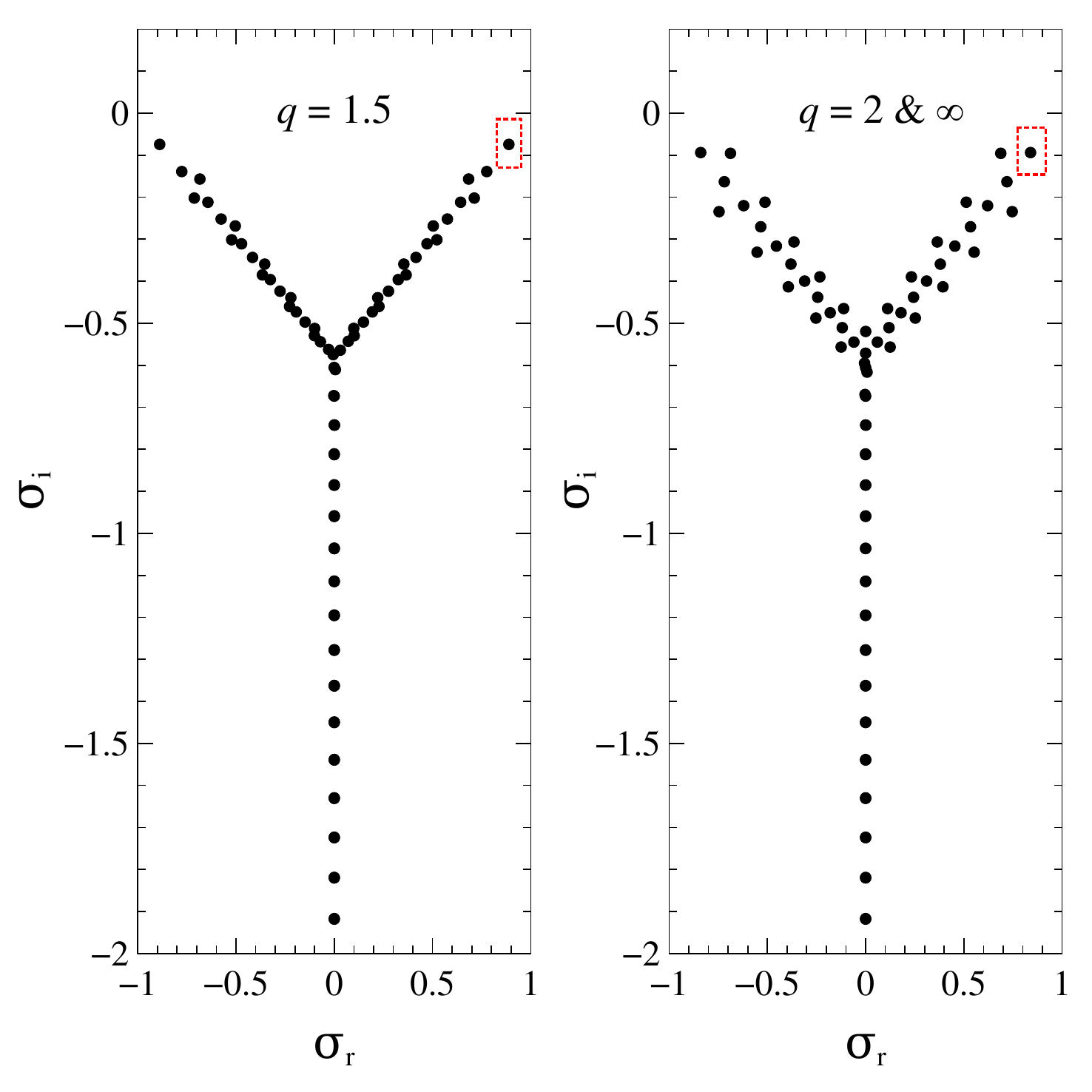}
\caption{Variation of $\sigma_i$ with $\sigma_r$ for $Re = 2000$ and $k_y=k_z = 1$ for the Keplerian flow ($q=1.5$), 
constant angular momentum flow ($q=2$) and plane Couette flow ($q\rightarrow\infty$). The latter two eigenspectra 
are identical. The dotted box represents the least stable mode for the respective cases.}
\label{fig:eval_three_cases}
\end{center}
\end{figure}

In the evolution of linear perturbation, let the linear solutions be
\begin{eqnarray}
 u = \hat u(x,t) e^{i\textbf{k}\cdot \textbf{r}},\\ 
 \zeta = \hat{\zeta}(x,t) e^{i\textbf{k}\cdot \textbf{r}},
\end{eqnarray}
with $\textbf{k} = (0, k_y, k_z)\ {\rm and}\ \textbf{r} = (0, y, z).$
Substitute these in equations (\ref{eq:orr_sommerfeld_eq}) and (\ref{eq:squire_eq}), neglecting non-linear terms, we 
obtain
\begin{eqnarray}
 \begin{split}
  \frac{\partial}{\partial t}Q +i\mathcal{L}Q +\Gamma = 0,
  \label{eq:lin_Q_eq}
 \end{split}
\end{eqnarray}
where
\begin{eqnarray}
\begin{split}
 Q = \begin{pmatrix}\hat{u}\\ \hat{\zeta}\end{pmatrix}, \ 
%\end{equation}
%\begin{equation}
 \mathcal{L} = \begin{pmatrix}
                \mathcal{L}_{11}& \mathcal{L}_{12}\\
                \mathcal{L}_{21}& \mathcal{L}_{22}
               \end{pmatrix},
\label{eq:the_L_matrix}\
\mathcal{D} = \frac{\partial}{\partial x},
\end{split}
\end{eqnarray}
\begin{eqnarray*}
\begin{split}
 \mathcal{L}_{11} =& (\mathcal{D}^2 - 
k^2)^{-1}\Bigl[k_yU\left(\mathcal{D}^2-k^2\right)-k_yU''-\frac{1}{iRe}\left(\mathcal{D}^2-k^2\right)^2\Bigr],\\
 \mathcal{L}_{12} =& \left(\mathcal{D}^2-k^2\right)^{-1}\frac{2k_z}{q},\\
 \mathcal{L}_{21} =& -\left(U'+\frac{2}{q}\right)k_z,\\
 \mathcal{L}_{22} =& k_yU-\frac{1}{iRe}\left(\mathcal{D}^2-k^2\right),
\end{split}
\end{eqnarray*}
and
\begin{equation}
 \Gamma = e^{-i\textbf{k}\cdot \textbf{r}}\begin{pmatrix}
           (\mathcal{D}^2 - k^2)^{-1}\Gamma_1\\
           \Gamma_2
          \end{pmatrix}.
          \label{eq:Gamma}
\end{equation}
Let us subsequently assume the trial solution of the equation (\ref{eq:lin_Q_eq}) be
\begin{equation}
 Q = AQ_xe^{-i\sigma t}-\frac{1}{\mathcal{D}_t+i\mathcal{L}}\Gamma,
 \label{eq:lin_trial_sol}
\end{equation}
where $\sigma$ is the eigenvalue corresponding to the particular mode and it is complex having real ($\sigma_r$) and 
imaginary ($\sigma_i$) parts,
 \begin{equation}
  Q_x = \begin{pmatrix}\phi^u(x)\\ \phi^{\zeta}(x)\end{pmatrix}
  \label{eq:Q_x}
 \end{equation}
and $\mathcal{D}_t$ stands for ${\partial}/{\partial t}.$ 
$Q_x$ is the eigenfunction corresponding to the homogeneous part of the equation (\ref{eq:lin_Q_eq}), i.e., $Q_x$ 
satisfies $\mathcal{L}Q_x = \sigma Q_x$. The first term of the right-hand side of the equation (\ref{eq:lin_trial_sol}) 
is due to the homogeneous part of the equation (\ref{eq:lin_Q_eq}), and the second term is due to the inhomogeneous part, 
i.e., the presence of $\Gamma$, of the same equation. Hence, $Q$ is influenced by the force $\Gamma$. However, the 
typical eigenspectra for the Keplerian flow ($q=1.5$), constant angular momentum flow ($q=2$) and plane Couette flow 
($q\rightarrow\infty$) for $Re = 2000$ and $k_y = k_z = 1$ are shown in the Fig.~\ref{fig:eval_three_cases}. Since 
$\mathcal{L}_{12}$ and $\mathcal{L}_{21}$ in the equation (\ref{eq:the_L_matrix}) are zero for the plane Couette flow 
and constant angular momentum flow, respectively, we obtain the same eigenspectra for both plane Couette, and constant 
angular momentum flows. We perform the whole analysis for the least stable modes for the respective flows, and these 
least stable modes are shown in the dotted box in Fig.~\ref{fig:eval_three_cases}.

%###########################################################################################%

\subsection{Nonlinear theory}
\label{sec:Non-linear theory}
For the non-linear solution, following similar work but in the absence of 
force \cite{Ellingsen_1970, Schmid_2002, Rajesh_2011}, we assume the 
series solution for velocity and vorticity, i.e. 
\begin{eqnarray}
 \psi = \sum_{n\rightarrow -\infty}^{\infty} \psi_n = \sum_{n\rightarrow -\infty}^{\infty} 
\bar{\psi}_n(t,x)e^{in(\textbf{k}\cdot 
\textbf{r} -\sigma_r t)},
\end{eqnarray}
when obviously $\bar{\psi}_{-n} = \bar{\psi}^*_n$ and $\psi$ could be any one of $u$ and $\zeta$. 

We substitute these in equations (\ref{eq:orr_sommerfeld_eq}) and (\ref{eq:squire_eq}) and collect the coefficients of the 
term $e^{i(\textbf{k}\cdot\textbf{r}-\sigma_r t)}$, to capture least nonlinear effect following, e.g., 
\cite{Ellingsen_1970, Rajesh_2011, Ghosh_2020}, from both sides and obtain

\begin{equation}
 \frac{\partial Q_1}{\partial t} -i\sigma_r Q_1 + i\mathcal{L}Q_1 = NL_1,
 \label{eq:Q_1_nonlin_eq_time_evo}
\end{equation}
where $NL_1 = \begin{pmatrix}\left(\mathcal{D}^2-k^2\right)^{-1}NL^{u}_1\\ NL^{\zeta}_1\end{pmatrix}$ and
 $Q_1 = \begin{pmatrix}\bar{u}_1(x,t)\\ \bar{\zeta}_1(x,t)
 \end{pmatrix}$.
We assume the solution for $Q_1$ to be
\begin{equation}
 Q_1 = \sum_{m=1}^{\infty}{A_{t,m}} {Q_{x,m}} - \frac{1}{\mathcal{D}_t+i\mathcal{L}}\Gamma,
 \label{eq:Q_1_total_trial_sol}
\end{equation}
where $m$ stands for various eigenmodes.

However, to the first approximation, our interest is in the least stable mode. Similar descriptions in two dimensions 
\cite{Ellingsen_1970} without $\Gamma$ and three-dimensional Keplerian disks\cite{Rajesh_2011} without $\Gamma$ are 
already there in the in the literature and we mostly follow their formalism. We, therefore, omit the summation, and 
subscript $m$ in the equation (\ref{eq:Q_1_total_trial_sol}), and 
obtain
\begin{equation}
 Q_1 = {A_{t}} {Q_{x}} - \frac{1}{\mathcal{D}_t+i\mathcal{L}}\Gamma.
 \label{eq:Q_1_trial_sol}
\end{equation}
We then substitute the equation (\ref{eq:Q_1_trial_sol}) in the equation (\ref{eq:Q_1_nonlin_eq_time_evo}) and using 
bi-orthonormality between $Q_x$ and its conjugate function $\tilde{Q}_x$, we obtain

\begin{eqnarray}
\frac{dA_t}{dt} - \sigma_i A_t +\mathcal{N} = p|A_t|^2A_t, 
\label{eq:mod_landau_eq}
\end{eqnarray}
where
\begin{eqnarray}
\mathcal{N} = \int_{-1}^{1}{dx \tilde{Q}_x^{\dagger}\Gamma'}
\label{eq:N}
\end{eqnarray}
and 
\begin{eqnarray}
p = \int_{-1}^{1}{dx \tilde{Q}_x^{\dagger}\mathcal{S}},
\end{eqnarray}
where 
\begin{eqnarray}
\begin{split}
\Gamma' =&-\Gamma + i\sigma_r\Bigl(\frac{1}{\mathcal{D}_t+i\mathcal{L}}\Bigr)\Gamma& \\&= - \Gamma + 
i\sigma_r(t-i\mathcal{L}t^2)(1+\mathcal{L}^2t^2)^{-1}\Gamma,&
\label{eq:exp_for_Gamma'}
\end{split}
\end{eqnarray}
$\mathcal{S}$ is the spatial contribution from the nonlinear terms. For the greater details, see Ghosh and 
Mukhopadhyay\cite{Ghosh_2020}.

Throughout the paper, $\Gamma$ from the equation (\ref{eq:Gamma}) has been decomposed as  
$\Gamma \rightarrow \Gamma \begin{pmatrix}
          1\\
           1
          \end{pmatrix}$ by adjusting $\Gamma_1$ and $\Gamma_2$, as they are only the free parameters.

\begin{figure}
\begin{center}
\includegraphics[width=3.5in]{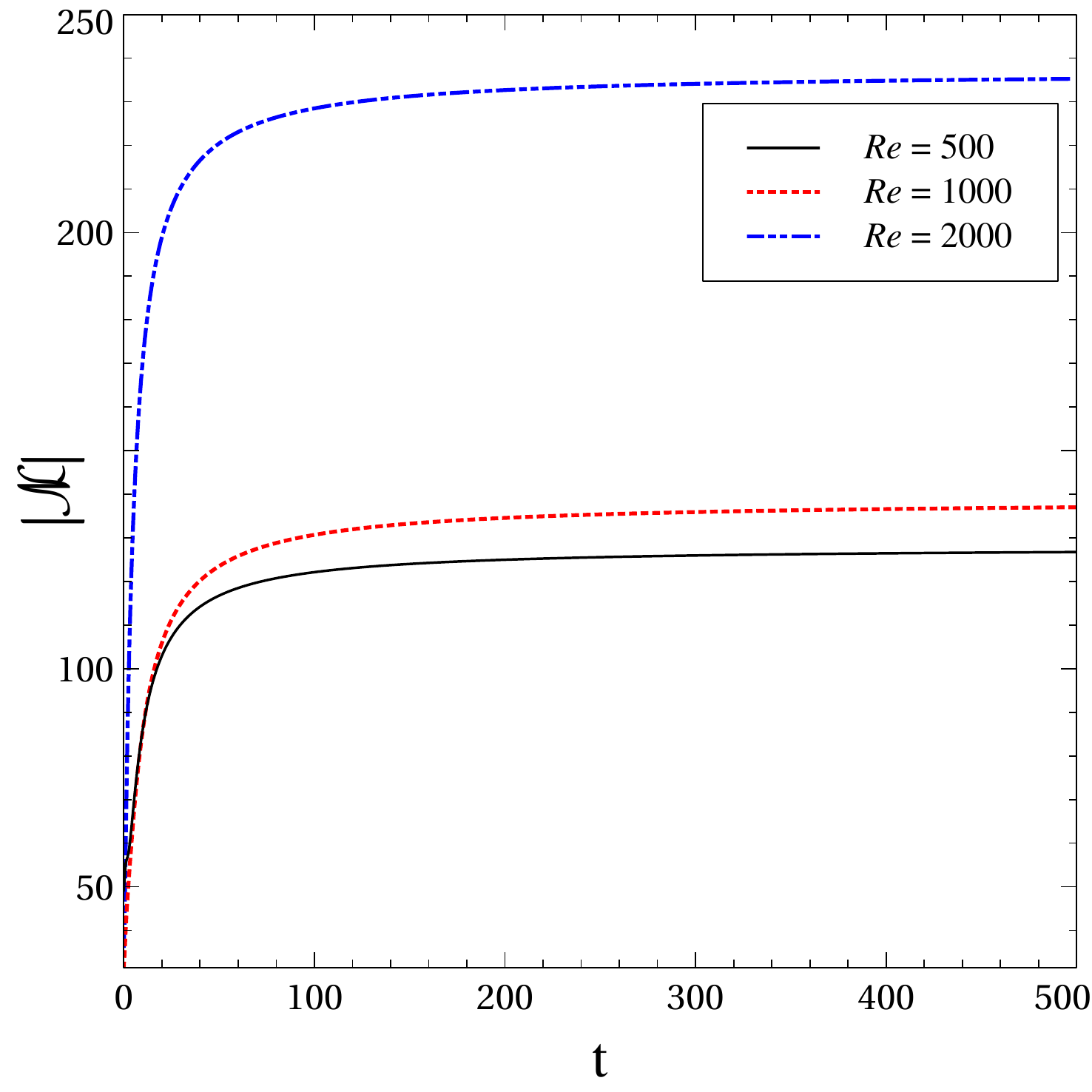}
\caption{Variation of $\mathcal{N}$ as a function of $t$ for $Re=500, \ 1000$ and 2000, for $\Gamma = 10^4$ and 
$k_y=k_z=1$, corresponding to the respective least stable modes.}
\label{fig:N_vs_t_R_500_1000_2000}
\end{center}
\end{figure}
The evolution of $\mathcal{N}$ has been shown in Fig.~\ref{fig:N_vs_t_R_500_1000_2000} for the parameters mentioned in 
the figure. It shows that $\mathcal{N}$ becomes saturated beyond a certain time, and the saturation depends on $Re$. As 
$Re$ increases, the saturation increases. 
\begin{figure}
\begin{center}
\includegraphics[width=3.5in]{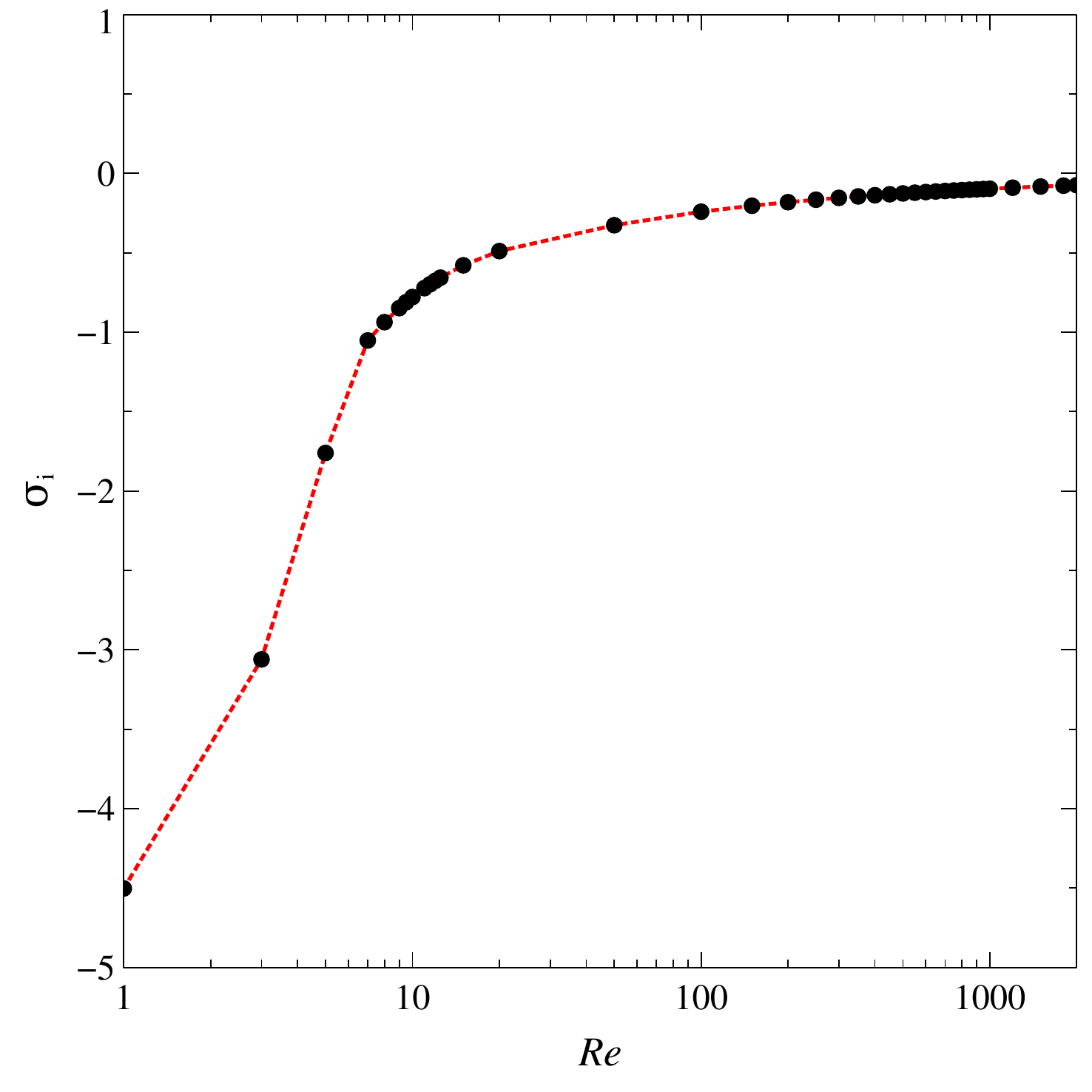}
\caption{Variation of the least stable $\sigma_i$ as a function of $Re$ for $k_y = k_z=1$ in the Keplerian flow.}
\label{fig:R_vs_sigma_i_kep}
\end{center}
\end{figure}
Fig.~\ref{fig:R_vs_sigma_i_kep} shows the variation of the least stable (the one in the dotted box in 
Fig.~\ref{fig:eval_three_cases}) as a function of $Re$ in the case of the Keplerian flow. It tells us that the least 
stable mode approaches zero as $Re$ is increased. 
%####################################################################################

\section{Results}
\label{sec:Results}

Equation (\ref{eq:mod_landau_eq}), which is a nonlinear equation, tells us about the evolution of the amplitude of the 
perturbation. To explore the evolution of linear and nonlinear perturbations, we use the equation 
(\ref{eq:mod_landau_eq}) accordingly. For the evolution of the linear perturbation, we get rid of the nonlinear term in 
equation (\ref{eq:mod_landau_eq}). 

\subsection{Evolution of $|A|_t$ in the linear regime}
\label{sec:Evolution of $|A|_t$ in the linear regime}
In the linear regime, the equation (\ref{eq:mod_landau_eq}) becomes 
\begin{eqnarray}
\frac{dA_t}{dt} = \sigma_i A_t -\mathcal{N}.
\label{eq:mod_landau_eq_lin}
\end{eqnarray}
This equation tells that at large $t$, $|A_t|$ becomes $|\mathcal{N}|/|\sigma_i|$\cite{Ghosh_2020}.

\begin{figure}
\begin{center}
\includegraphics[width=5in]{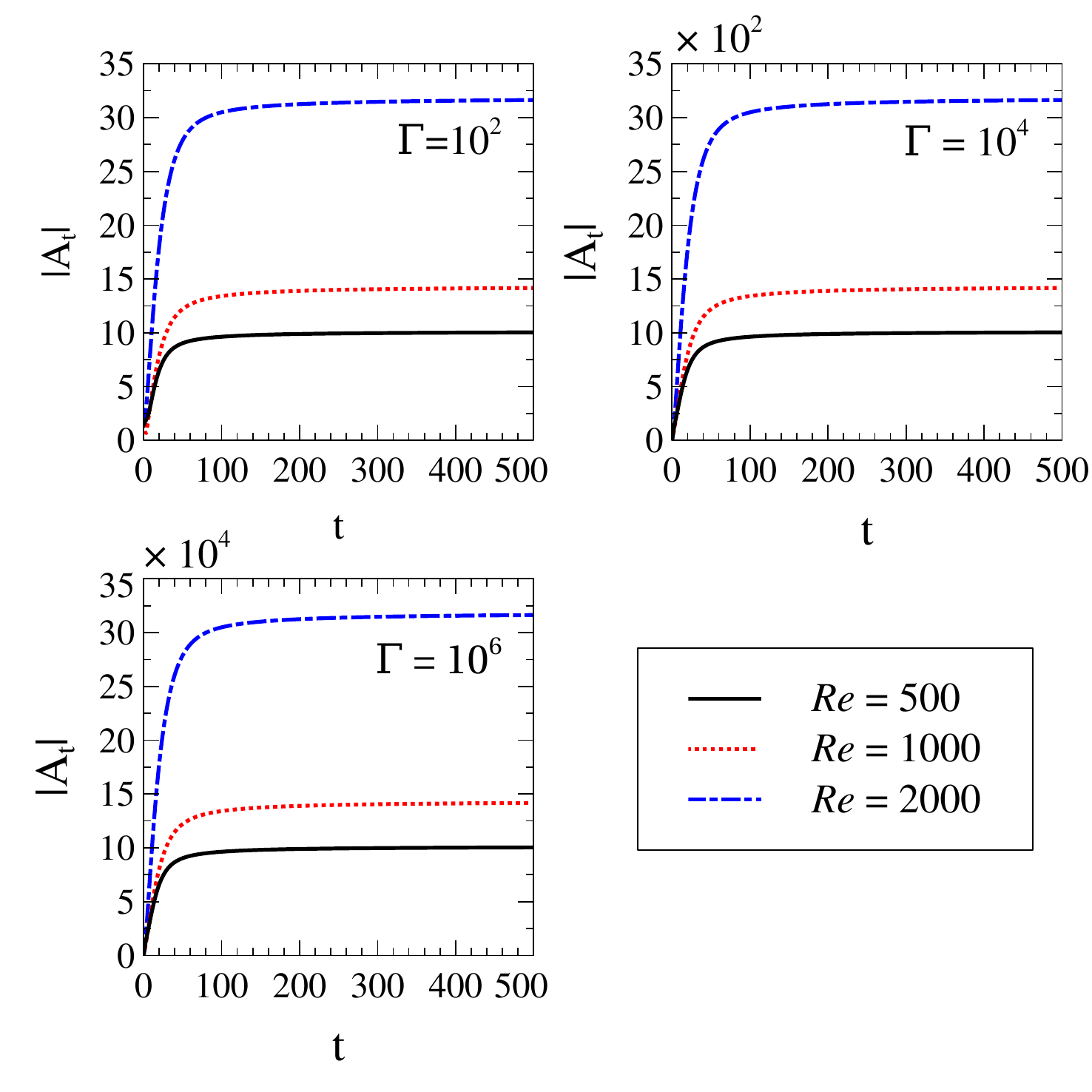}
\caption{Variation of $|A_t|$ as a function of $t$ for three sets of $Re$ and $\Gamma$ with $k_y = k_z=1$ for linear 
analysis in the Keplerian flow ($q=1.5$).}
\label{fig:A_t_vs_t_lin_3_diff_R_3_diff_gamma}
\end{center}
\end{figure}

Fig.~\ref{fig:A_t_vs_t_lin_3_diff_R_3_diff_gamma} shows the variation of $|A_t|$ as a function of $t$ for various values 
of $Re$ and 
$\Gamma$. Fig.~\ref{fig:A_t_vs_t_lin_3_diff_R_3_diff_gamma} also suggests the scaling relation between saturated 
$|A_t|$ and $\Gamma$ to be
\begin{eqnarray}
 |A_t|\propto \Gamma
 \label{eq:A_t_Gamma}
\end{eqnarray}
for a fixed $Re$. Apart from that, we notice that for a fixed $\Gamma$, $|A|_t$ becomes saturated at 
$|\mathcal{N}|/|\sigma_i|$ beyond a certain time, and the saturation increases as $Re$ increases. This saturation is 
independent of the initial condition\cite{Ghosh_2020}. This is because at a fixed $\Gamma$, $|\mathcal{N}|$ becomes 
saturated (see Fig.~\ref{fig:N_vs_t_R_500_1000_2000}) at a fixed value, and the saturated value increases as $Re$ 
increases, as well as $|\sigma_i|$ increases as $Re$ increases (see Fig.~\ref{fig:R_vs_sigma_i_kep}). Now, in 
the accretion disk, $Re$ is huge\cite{Mukhopadhyay_2013} ($\gtrsim10^{14}$). The smaller $\Gamma$, therefore, can bring 
nonlinearity in the accretion disk, as for huge $Re$, $|\sigma_i|$ is infinitesimally small. Hence, 
$|\mathcal{N}|/|\sigma_i|$ becomes enormous. This huge saturation of $|A_t|$ could bring nonlinearity and hence plausibly 
turbulence in the underlying Keplerian flow.

\subsection{Evolution of $|A_t|$ in the nonlinear regime}
\label{sec:Evolution of $|A_t|$ in the nonlinear regime}
\begin{figure}
\begin{center}
\includegraphics[width=5in]{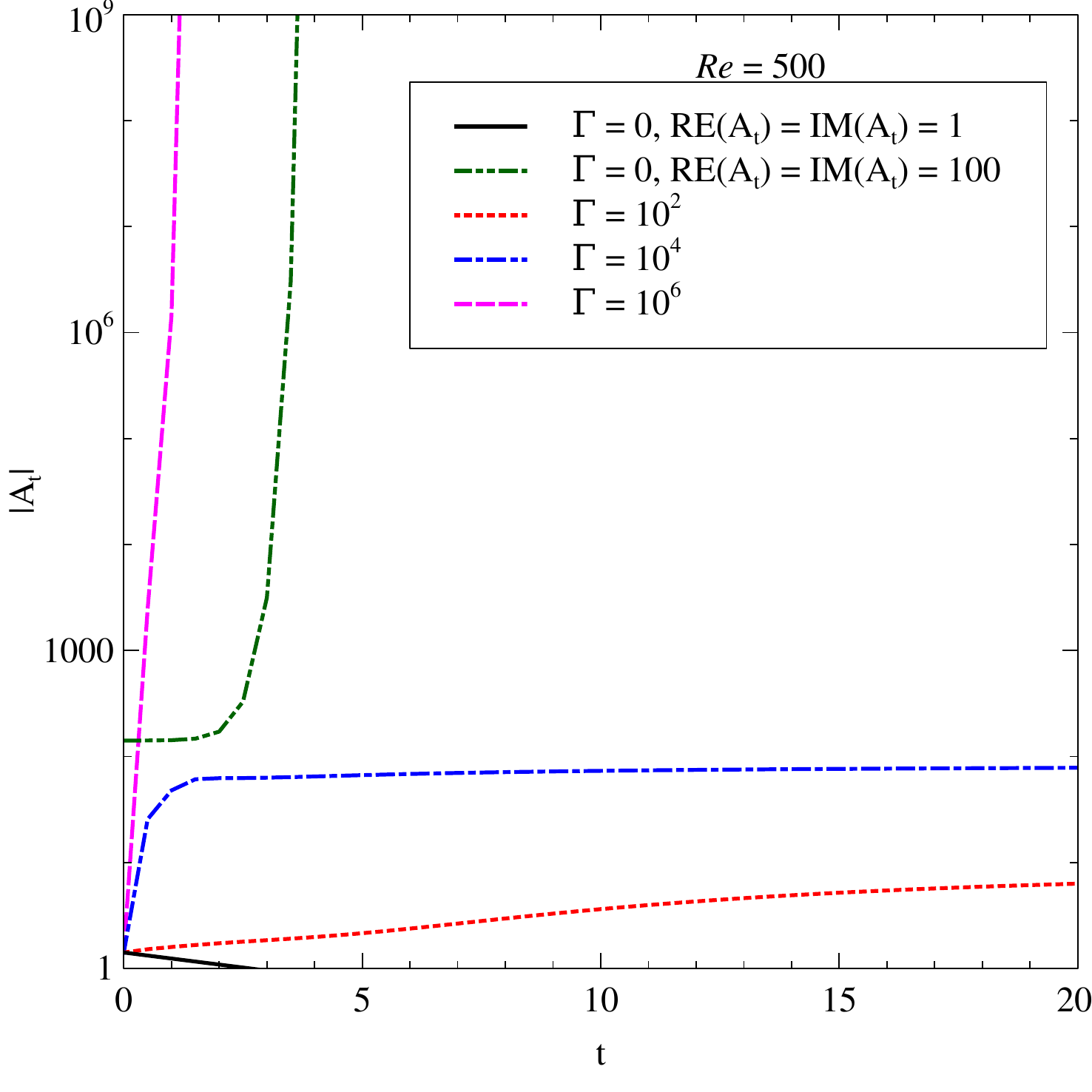}	
\caption{Variation of $|A_t|$ as a function of $t$ with $k_y = k_z=1$ for nonlinear analysis in the Keplerian flow for 
$Re=500$ and four different $\Gamma$. For $\Gamma\neq 0$, initial condition is RE($A_t$) = IM($A_t$) = 1.}
\label{fig:A_t_vs_t_nonlin_landau_modi_landau_diff_gamma_R_500}
\end{center}
\end{figure}

In the nonlinear regime, the evolution of $|A_t|$ is described by the equation (\ref{eq:mod_landau_eq}). 
Fig.~\ref{fig:A_t_vs_t_nonlin_landau_modi_landau_diff_gamma_R_500} describes the variation of $|A_t|$ as a function of 
$t$ for nonlinear perturbation in the Keplerian flow. In the figure, $RE(A_t)$ and $IM(A_t)$ indicate the real and 
imaginary parts of initial $A_t$, respectively. It tells us that for $\Gamma = 0$, depending on the initial amplitude, 
$|A_t|$ may diverge (green line) or $|A_t|$ may decay (continuous black line) to zero from the 
initial amplitude. When there is no noise, i.e., $\Gamma=0$, the diverging solution of $|A_t|$ occurs when the initial 
amplitude is beyond a critical amplitude\cite{Ghosh_2020}. It is apparent that the onset of the nonlinearity depends on 
the initial amplitude of perturbation in the absence of the force, but it does not depend on the same in the presence of 
force\cite{Ghosh_2020}. The divergence of $|A_t|$ and hence the onset of nonlinearity and plausible turbulence depends 
only on the strength of the force, as shown by dashed (magenta) line, compared to dot-dashed (blue) and dotted (red) 
lines, in Fig.~\ref{fig:A_t_vs_t_nonlin_landau_modi_landau_diff_gamma_R_500}. 
  
\section{Conclusion}
If there is no extra force involved in the system, then the equation (\ref{eq:mod_landau_eq}) becomes the usual Landau 
equation, which is
\begin{eqnarray}
 \frac{dA_t}{dt} = \sigma_i A_t + p|A_t|^2A_t,
 \label{eq:original_landau_eq}
\end{eqnarray}
which can be further recast to
 \begin{eqnarray}
\frac{d|A|^2}{dt}  = k_1 |A|^2 + k_2|A|^4,
\label{eq:landau_eq}
\end{eqnarray}
where $A$ is the amplitude of the nonlinear perturbations for the corresponding system, $k_1$ is 2$\sigma_i$ and 
$k_2$ is the real part of $2p$, i.e., 2$p_r$. Its solution is 
\begin{eqnarray}
 |A|^2 = \frac{A_0^2}{-\frac{k_2}{k_1}A_0^2+\left(1+\frac{k_2}{k_1}A_0^2\right)e^{-k_1t}},
 \label{eq:sol_landau_eq}
\end{eqnarray} where $A_0$ is the initial amplitude. Depending on the sign (positive/negative) of $k_1$ and $k_2$, there 
are four\cite{Drazin_2004, Schmid_2002} different possible evolutions of $|A|$. These given below.
\begin{itemize}
 \item In the present context of shear flows, $k_1$ (i.e., $\sigma_i$) is negative, but $k_2$ is positive. Therefore, 
there will be a threshold for initial amplitude $A_i$, determining the growth of perturbation. If the initial amplitude 
$A_0<A_i$, then 
\begin{eqnarray}
|A|^2\sim\frac{A_i^2A_0^2e^{k_1 t}}{A_i^2-A_0^2}
\end{eqnarray} 
at a large $t$. Therefore, $|A|^2\rightarrow 0$ for $A_0<A_i$ at $t\rightarrow\infty$. However, if $A_0>A_i$, then 
$|A|^2\rightarrow\infty$ at $t\rightarrow \ln(1-A_i^2/A_0^2)/k_1$.
\item If both $k_1$ and $k_2$ would be positive, $|A|^2$ blows up after a finite time. Hence, there will be a fast 
transition to turbulence.
\item On the other hand, if $k_1>0$ but $k_2<0$, then $|A|^2\rightarrow k_1/|k_2|$ at $t\rightarrow\infty$. In this case, 
$|A|^2$ at a large $t$ does not depend on $A_0$.
\item Obviously, for $k_1$ and $k_2$ both negative, $|A|^2$ decays fast.
\end{itemize}

However, we have shown that the saturation in $|A_t|$ is $|\mathcal{N}|/|\sigma_i|$ in the linear regime. We have 
also argued that depending on $Re$ and $\Gamma$, the system may already be in the nonlinear regime.  The evolution of 
$|A_t|$ at the linear regime in our case, i.e., with extra force, is similar to that of $|A|$ from the equation 
(\ref{eq:landau_eq}), i.e. without force, for $k_1>0$ and $k_2<0$. In the Keplerian flow, $k_1$, i.e. $\sigma_i$, is 
negative, 
but $k_2$, i.e. $p_r$, could be positive. In the presence of extra force, the Landau equation modifies in such a way 
that the solution in the linear regime itself mimics the Landau equation without force (i.e., equation 
\ref{eq:landau_eq}), however, with $k_1>0$ and $k_2<0$. Further, in the nonlinear regime, the amplitude $A_t$ (i.e., with 
extra force included) diverges beyond a certain time, depending on $Re$ and $\Gamma$. In the nonlinear regime, the Landau 
equation in the presence of extra force but negative $k_1$ ($\sigma_i$) is, therefore, mimicking the Landau equation 
without force but with positive $k_1$ and $k_2$. Essentially, the extra force effectively changes the sign of $k_1$ (i.e., 
$\sigma_i$) for the Landau equation without force. Speaking in another way, the very presence of extra force destabilizes 
the otherwise stable system. Thus, the presence of force plays an important role in developing nonlinearity and turbulence 
in the case of Rayleigh stable flows.

\section*{Acknowledgment}

S.G. acknowledges DST India for INSPIRE fellowship. This work is partly supported by a fund of Department of Science and 
Technology (DST-SERB) with research Grant No. DSTO/PPH/BMP/1946 (EMR/2017/001226).
\bibliographystyle{ws-procs961x669}
\bibliography{accretion_disks_jets}

\end{document}